\documentclass[10pt]{article}
\newcommand{\amean}[2]{\lfloor #1 \rfloor_{#2}}
\newcommand{\leftangle}{\langle}
\newcommand{\rightangle}{\rangle}
\usepackage{graphicx}
\usepackage{subfigure}
\usepackage{amsmath}
\newcommand{\Name}[1]{{\scshape #1}}
\newcommand{\Review}[1]{{\itshape #1}}
\newcommand{\Vol}[1]{%
   {\bfseries #1}
}
\newcommand{\Year}[1]{#1}
\newcommand{\Page}[1]{%
  {\normalfont #1}
}

\newcommand{\Book}[1]{{\itshape #1},}
\newcommand{\Publ}[1]{{\normalfont(#1)}}
\newcommand{\REVIEW}[4]{\Review{#1}, \Vol{#2} (\Year{#3}) \Page{#4}}

\begin{document}

\begin{center}
SAMPLE-DEPENDENT  PHASE TRANSITIONS IN DISORDERED EXCLUSION MODELS

\vskip10pt
C. Enaud
and B. Derrida
\vskip .5cm
Laboratoire de Physique Statistique \footnote{email:  enaud@lps.ens.fr
and derrida@lps.ens.fr}, \\
Ecole Normale Sup\'erieure, 24 rue Lhomond, 75005 Paris, France

\vskip20pt

\end{center}

\vskip20pt
\noindent {\bf Abstract}


We give numerical evidence that the location of the first order phase transition
between the low and the high density phases of the one dimensional
asymmetric simple exclusion process with open boundaries becomes
sample dependent when quenched disorder is introduced for the hopping
rates.
\newpage

The effect of quenched disorder on phase transitions is one
of the most studied  and best understood aspects of the theory of
disordered systems \cite{FROHLICH, STINCHCOMBE83, SHALAEV94,  HARRIS74,
STINCHCOMBE02, CARDY97, AIZENMAN89, IMRY75, BERCHE98}. For
equilibrium systems, the Harris criterion \cite{HARRIS74, STINCHCOMBE02} allows one to decide whether the critical behavior of
a second order phase transition is altered by a weak disorder. One also 
knows \cite{CARDY97, AIZENMAN89, IMRY75, BERCHE98} that first order transitions are
suppressed by a weak disorder in low enough dimension.

It is now well established that non equilibrium systems can exhibit phase
transitions even in one dimension (for a review see \cite{EVANS99}). For
\nocite{KRUG91}
simple systems as the asymmetric exclusion with open
boundaries (the TASEP), the phase diagram is known exactly
\cite{KRUG91, DERRIDA92, DERRIDA93JPA, SCHUTZ93, POPKOV99}. It is then a natural question to investigate what is the effect
of disorder on such a phase diagram \cite{KRUG99, STINCHCOMBE02}.

	The one dimensional asymmetric simple exclusion process (ASEP)\cite{ 
SPOHN91, SCHUTZ00, HINRICHSEN00} has been considered in several contexts ranging
from biopolymerisation and electrophoresis to the study of  car traffic
\cite{CHOWDHURY00}. It is
also related to other problems of statistical physics like  surface growth and directed
polymers in  random media \cite{ZHANG95, KRUG97}.  It describes a one dimensional lattice of $L$ sites,  each
site of which is either empty or occupied by a single  particle
(exclusion rule). During each infinitesimal time
interval $dt$, a particle at site $i$ attempts to move to its right
neighboring site 
with probability $p dt$ and the move is successful only if the target
site $i+1$
is empty. At the boundary (site $1$ and site $L$) the dynamics is
modified as follows (open boundary conditions): during
 each infinitesimal time interval $dt$, a particle is
injected on site $1$  of the lattice with probability $\alpha
dt$ if this site is empty; and if a particle is present on
site $L$, it is
removed with probability $\beta dt$.
When the boundary parameters $\alpha$ and $\beta$ are varied, the model
exhibits several phase transitions \cite{KRUG91,DERRIDA92,SCHUTZ93,
DERRIDA93JPA,  POPKOV99}: second order phase
transitions along the lines $\alpha=p/2 <\beta$ (from a low density phase
to a maximal current phase) and $\beta=p/2<\alpha$ (from a high density
phase to a maximal current phase), and a first order transition from the
low to the high density phase along the line 
\begin{equation}\label{trans}
\alpha=\beta <\frac{p}{2}
\end{equation}

The goal of the present work is to study the effect of quenched disorder
on this first order phase transition.
Different ways of introducing quenched disorder have been considered for the ASEP. The most studied
is the \emph{particlewise} disorder \cite{KRUG96, KRUG99,CSAHOK94}. It
corresponds to the case where the
hopping rates $p$ depend on the particle which attempts to jump.
One can think of it as  a model of traffic along  a one lane road where each
vehicle has a different, random velocity. This model can be
exactly solved through a mapping to the zero-range process \cite{KRUG96}.

We consider here the less studied \emph{sitewise} disorder,
where to each lattice site is associated a random, quenched hopping rate
$p_i$. In
the traffic analogy, this would correspond to the case where localized
accidents or bottlenecks locally decrease the mean speed of vehicles.
Even the adjunction of a single localized defect in the hopping rate
$p_i$ transforms the ASEP into a difficult, and yet
unsolved problem  \cite{SCHUTZ93wb, JANOWSKY94,  KOLOMEISKY98}.  There are
only few analytical results in the case of site disorder, the most
noticeable
ones being the
existence of an hydrodynamic limit for the density, the concavity of the current of particles as a function of the density of particles \cite{SEPPALAINEN99} and
the reflection invariance of the current \cite{GOLDSTEIN98}.
Monte-Carlo simulations on ASEP  on a
ring geometry (i.e. with periodic boundary conditions)\cite{TRIPATHY98,BENGRINE98} have shown that 
the current-density relation  is modified by quenched sitewise disorder,
with the appearance of a plateau between two densities $\rho_c$
and $1-\rho_c$  where the current becomes independent of the density of
particle $\rho$ on the ring. Tripathy and Barma \cite{TRIPATHY98} have also shown
that this plateau corresponds to a \emph{segregated-density regime} where
the disorder induces a phase separation 
 between a high density  and a
low density phase. Outside this plateau (i.e. for densities $\rho<\rho_c$
or $\rho > 1-\rho_c$), the steady state profile  is  uniform on a
macroscopic scale 
(but with  microscopic variations).
Some bounds for $\rho_c$ have been obtained \cite{KRUG99}.
When the distribution of the  $p_i$ allows values as close as possible to
$0$ a third regime occurs, in which the steady state current vanishes in the
thermodynamic limit \cite{KRUG99, TRIPATHY98} .

In the open boundaries case, when particles are injected at site $1$ and
removed at site $L$ on the last one, one expects the system to  have the
same
3 phases  (low density, high density and maximal
current) as in the pure case.
The question then is how the nature and the location of the phase
transitions are modified by the effect of disorder.

\bigskip
Here we investigate numerically the effect of disorder on the first order
transition (\ref{trans}).
We choose a binary distribution of the bulk hopping rates  $p_i$.
\begin{align}\label{distp}
 p_i&=\left\{
\begin{array}{c}
p_\text{min} \text{ with probability } \frac{1}{2}\\
p_\text{max} \text{ with probability } \frac{1}{2}
\end{array}
\right.
\end{align}
with $p_\text{min}=0.8$, $p_\text{max}=1.2$.

\bigskip
In equilibrium systems (with short range inteactions), the properties of finite systems depend smoothly on external parameters and phase transitions usually occur in the
thermodynamic limit, i.e. in the limit of an infinite system size. 
The ASEP on a finite chain with or without disorder is a Markov process
with a stationnary state which also depends smoothly on the parameters
$\alpha$ and $\beta$ and one needs to take the limit of an infinite size to observe phase transitions.

As the $p_i$'s depend on $i$, one needs to choose a procedure to add new sites to a given sample in order to take the thermodynamic limit.
In the following we compare two different procedures: 
\begin{itemize}
\item Procedure C (C for center): we increase the size of a given sample by adding new sites at the center of the sample. So as $L$ increases, the $p_i$ close to the boundaries remain unchanged and sites with new $p_i$'s are introduced at the center, i.e. further and further from the boundaries
\item Procedure B (B for boundary): we add the new sites at the two boundaries, whereas the center of the sample remains unchanged.
\end{itemize}
\begin{figure}
\setlength{\abovecaptionskip}{3pt}
\setlength{\belowcaptionskip}{0pt}
\setlength{\floatsep}{2pt plus 1pt minus 2pt}
\centering{\subfigure[\label{rhogrowcenter}]{\includegraphics[width=.40\textwidth,trim=0 5 0 0 ]{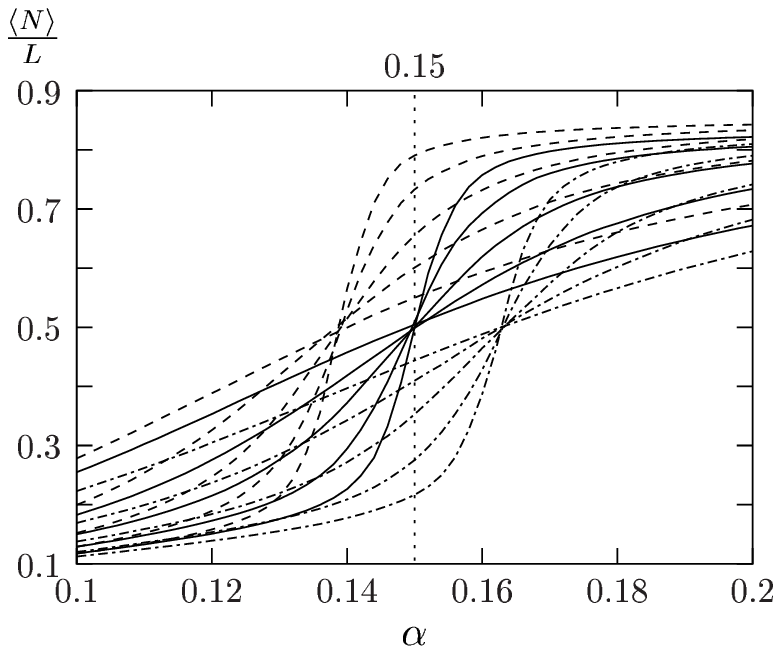}}\hfill
\subfigure[ \label{rhogrowbound}]{\includegraphics[width=.40\textwidth,trim=0 5 0 0 ]{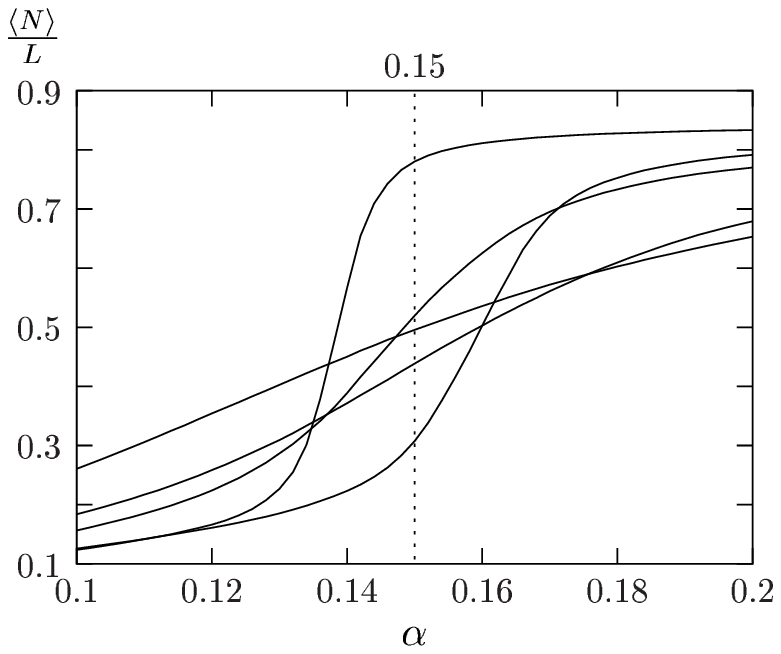}}\\
\subfigure[ \label{rhopure}]{\includegraphics[width=.40\textwidth,trim=0 5 0 20 ]{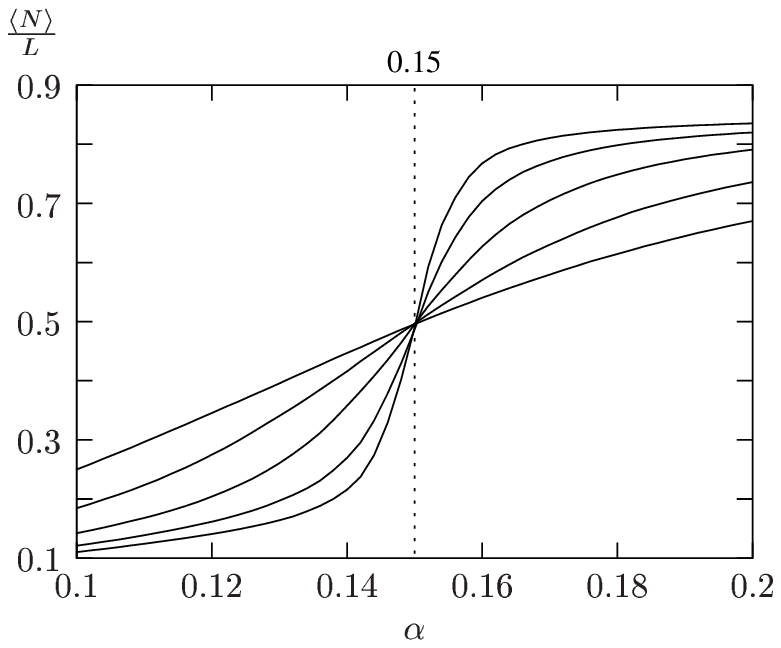}}}
\caption{ Mean density $\frac{\langle N \rangle}{L} $ versus $\alpha$  for  a sample
of size 11, 21, 41, 81 and 161.
In figure \ref{rhogrowcenter}, the system size is increased by adding new sites at the center (procedure C), and 3 samples are shown. In figure \ref{rhogrowbound}, the size is increased by adding new sites on the sides (procedure B) and a single sample is shown. In figure \ref{rhopure}, the pure case (all the $p_i=1$) is shown for comparison. The dotted line shows on the three figures the transition point of the pure TASEP.
}\label{rho}
\end{figure}

For each sample of size $L$, we make  Monte Carlo simulations for
different values of $\alpha$ at fixed
$\beta=0.15$ and different values of $\alpha$.
Figure \ref{rho}  shows the steady state density
$\langle\frac{N}{L}\rangle$ where  $N$ is the total number of particles in the system
as a function of $\alpha$. The density is averaged  over
typicaly $3\times 10^5$ updates per site. Figure \ref{rhogrowcenter} corresponds to
three different samples when the size is increased according to procedure (C); figure 
\ref{rhogrowbound} to a single sample when the size is increased by the
sides (procedure (B)),
and figure \ref{rhopure} shows the pure case ($p_i=1$).
As $\alpha$  varies, one sees a transition from a low density for small
$\alpha$ to a high density for large $\alpha$. The slope at the crossing point becomes steeper
and steeper  when the size increases; the position of the value of
$\alpha$ where the slope is maximum  is
sample dependent but size independent in figure \ref{rhogrowcenter} when the size is increased
by the center. On the contrary it is size dependent in figure \ref{rhogrowbound} when the
size is increased by the sides.

\begin{figure}
\setlength{\abovecaptionskip}{3pt}
\setlength{\belowcaptionskip}{0pt}
\setlength{\floatsep}{2pt plus 1pt minus 2pt}
\centering{\subfigure[
\label{sigmagrowcenter}]{\includegraphics[width=.40\textwidth,trim=0 5 0 0 ]{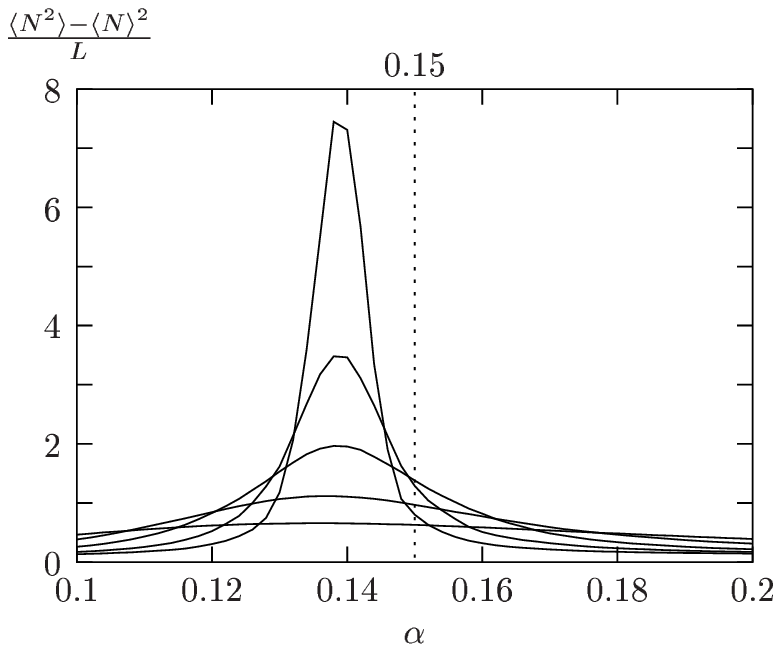}}\hfill
\subfigure[ \label{sigmagrowbound}]{\includegraphics[width=.40\textwidth,trim=0 5 0 0 ]{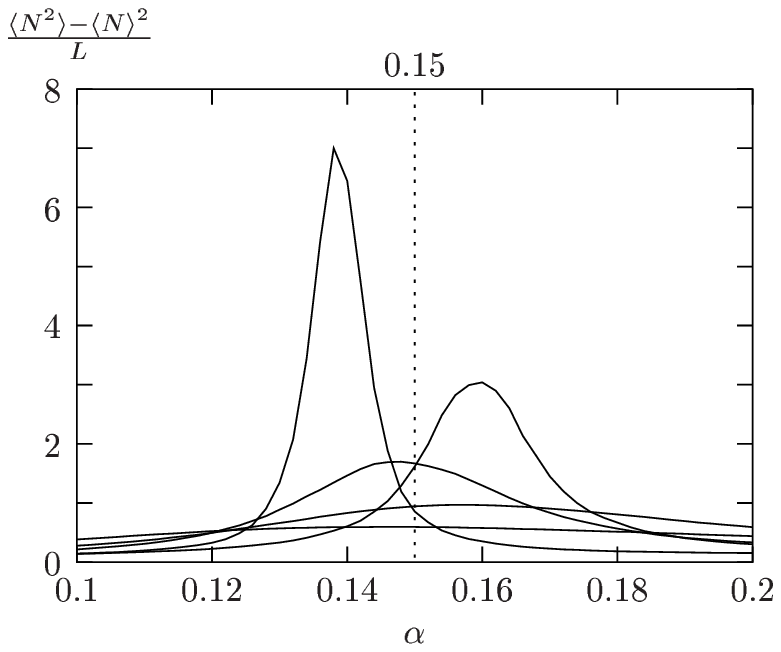}}\\
\subfigure[ \label{sigmapure}]{\includegraphics[width=.40\textwidth,trim=0 5 0 20 ]{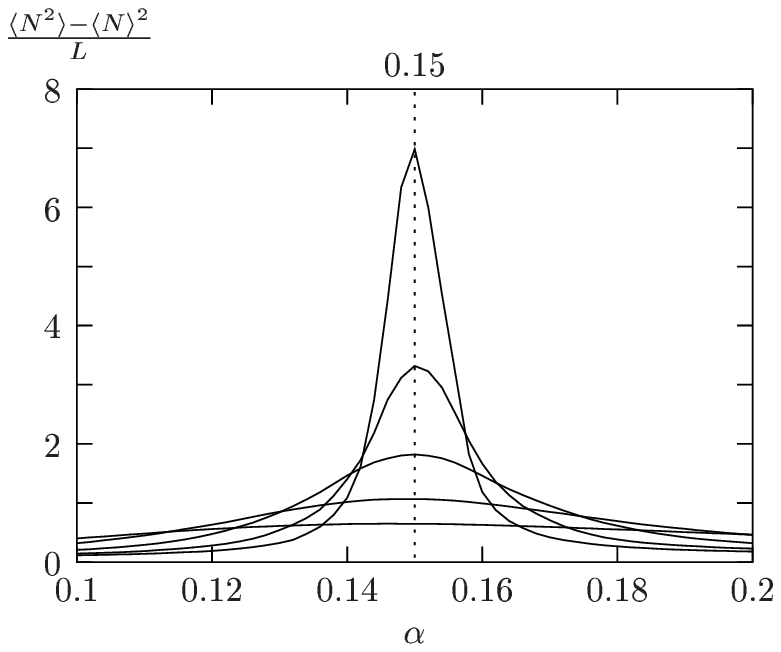}}} \caption{ Variance  $(\leftangle N ^2\rightangle-\leftangle
N \rightangle ^2)/L$ versus $\alpha$  for a single sample
of size 11, 21, 41, 81 and 161.
In figure \ref{sigmagrowcenter}, the system size is increased by the center (procedure C). In figure \ref{sigmagrowbound}, the size is increased by the sides (procedure B). In figure \ref{sigmapure}, the pure case (all the $p_i=1$) is shown for comparison. The dotted line shows the position of the transition point for the pure TASEP.
}\label{sigma}
\end{figure}
In the same way, we show in figures (\ref{sigma})  the variance $(\langle N^2 \rangle
- \langle N\rangle^2)/L$ as the size is increased.  
 In
the pure case (figure \ref{sigmapure}), the location of the maximum of the curve
converges to $\alpha_c=\beta$,  the phase transition
point of the infinite system as the size  increases. When the size of the
disordered  samples
is increased by the center (figure
\ref{sigmagrowcenter}), the location of the maximum of the curve
converges in a way similar to the pure  case but to a critical value  for
$\alpha$ which depends on the initial sample. This suggests
that in the thermodynamic limit, the system undergoes a
phase transition for a critical value $\alpha_c$ which is sample
dependent and depends on the
hopping rates near boundaries. Moreover, as the size dependence of the
maximum of the curve is the same in figure \ref{rhogrowcenter} and
\ref{rhopure}, one expects that the transition remains first order
in presence of disorder.\\
When the size of the sample is increased by the sides (figure
\ref{sigmagrowbound}), the location of the maximum does not converge in
the thermodynamic limit.

\bigskip

\begin{figure}
\centering{\includegraphics[width=0.70\textwidth]{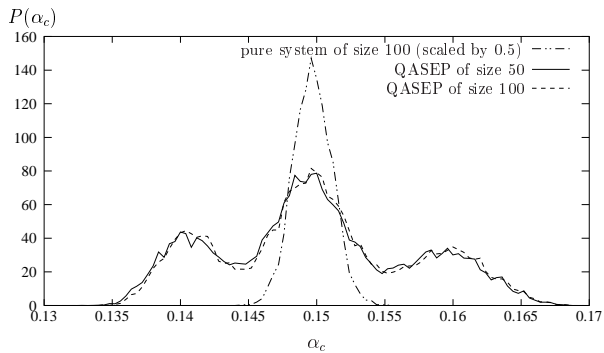}}
\caption{Distribution $P(\alpha_c)$ of the pseudo transition point $\alpha_c$ for $\beta=0.15$
when the hopping rates $p_i$ are given by (\ref{distp}) (full line: size
$L=50$; dashed line: size $L=100$). The distribution does not shrink as
the size increases. The distribution (divided by 2) of the pseudo transition point for the pure system measured via the
same method is shown for comparison.\label{alphacdens}}
\end{figure}

Figure \ref{alphacdens} shows the distribution of the pseudo transition point  (the value of $\alpha$ for which the variance in figure \ref{sigma} is maximum)
 for quenched systems of size $L=50$ and $L=100$. We generated $40000$
samples for each size and each sample was simulated  over $300000$ steps
per site, after a sufficiently long transient time to reach the steady
state. The histogram is
then obtained by discretizing the $\alpha_c$-axe in boxes of length
$4\times 10^{-4}$.  
In order to estimate the numerical noise on the determination of the
critical-point coming from the finite time length ($3 \times 10^5$) of our simulation, we include the distribution
obtained over $20000$ samples of the pure system (the distribution for the pure case has been divided by $2$ to make the top of the peak visible in figure \ref{alphacdens}).

The distribution in presence of disorder remains unchanged  as the system size $L$
increases by a factor $2$, which suggests that the same distribution
would be observed in the thermodynamic limit.
As the distribution $P(\alpha_c)$ in presence of disorder is much broader than in the pure case,
 its width is not due to the limited time of the measurement. Still, all the details of 
the distribution $P(\alpha_c)$ cannot be resolved on a scale smaller than
the width of the $P(\alpha_c)$ of the pure system. Therefore longer
simulations should reveal more structure at smaller scales of the distribution $P(\alpha_c)$ (while the distribution for the pure case should become narrower).

 \begin{figure}
\centering{\includegraphics[width=.70\textwidth]{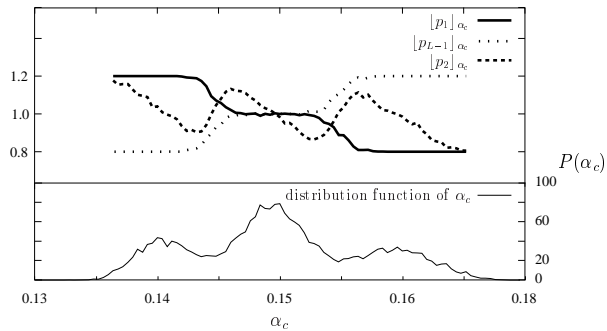}
\caption{$\amean{p_1}{\alpha_c}$, 
$\amean{p_2}{\alpha_c}$ and  $\amean{p_{L-1}}{\alpha_c}$ versus the position of the
critical point $\alpha_c$. The full curve is the same distribution of the
critical point $\alpha_c$ as in figure \ref{alphacdens}. We see that all
samples which have a pseudo transition point $\alpha_c$ in the left peak
have $p_1=p_{max}$ and $p_{L-1}=p_{min}$ and all the sample which have a
pseudo transition point $\alpha_c$ in the right peak have $p_1=p_{min}$ and
$p_{L-1}=p_{max}$.
\label{pmean1}}}
\end{figure}

Figure \ref{pmean1} tries to evaluate the correlation between the value of
$\alpha_c$ and the hopping rates close to the boundaries.
We note by $\amean{.}{\alpha_c}$ the average over the samples that
have a critical point $\alpha_c$ (up to the discretization of the
$\alpha_c$-axis).
On figures (\ref{pmean1}) we show the average of some hopping rates
$\amean{p_i}{\alpha_c}$, namely the first $p_1$, the second $p_2$ and the
last $p_{L-1}$, as a function of the
value of the critical point $\alpha_c$.
One sees that the three  maxima of the distribution of the critical
points can be characterized by the value of the first hopping rate $p_1$ and
$p_{L-1}$: the left peak corresponds to
$p_1=p_\text{max}$, $p_{L-1}=p_\text{min}$, the right peak to $p_1=p_\text{min}$, $p_{L-1}=p_\text{max}$.
The curve of the average of the second hopping rate
$\amean{p_2}{\alpha_c}$ as a function of $\alpha_c$ supports the
idea that the value of $p_2$ has a weaker influence on $\alpha_c$ than
$p_1$ or $p_{L-1}$. Therefore the picture which emerges is that
$\alpha_c$ is sample dependent and the influence of the $p_i$'s on
$\alpha_c$ decreases with their distance to the boundary (which could
lead to a fractal $P(\alpha_c)$ (smoothed out in our figure
\ref{alphacdens} due to the finite time length of our simulation)). The
fact that $\alpha_c$ is strongly correlated  to the $p_i$ close to 
the boundaries explains why $\alpha_c$ converges when procedure C is used
as in figure \ref{rhogrowcenter} and why it wanders around when procedure
B is used.

\bigskip

From a technical point of view, let us mention that, for each sample, we
made the measurement for all the values of $\alpha$ in a single simulation.
To do so, we introduced $K$  different classes of particles
\cite{ANDJEL88, FERRARI91}, indexed from $1$ to
 $K$, and whose dynamics is the same as in the TASEP, except that the particles of a higher class $K$
behave like holes in front
of particles of lower class $K'$ (when $K'<K$).
More precisely, we still impose the
exclusion rule, so there is no more than one particle on each site of
the system; and at each infinitesimal time $dt$, a particle on site $i$
attempts to jump to its right neighbor with probability $p_i dt$. The
jump succeeds only if the target site $i+1$ is empty or contains a particle of a
class higher than the particle which attempts the jump. If the jump
succeeds, the particles on site $i$ and $i+1$ are exchanged. \\
To simulate $K$ values $\alpha_1<\alpha_2<\cdots<\alpha_K$ of $\alpha$ the rule is to introduce particles of class $k$ at rate $\alpha_k-\alpha_{k-1}$ (with $\alpha_0=0$) at the left boundary provided that the first site is empty or occupied by a particle of higher class. At the right boundary, a particle present at site $L$ is removed at rate $\beta$ irrespective of its class.\\
To do our measurement of the number of particles for a given choice of
$\alpha_k$, we count as particles all particles of class $\leq k$ and as
holes all particles of class $\geq k+1$.

In this work, we have given numerical evidence that  for the ASEP
with open boundaries,
\emph{the location of the first order phase transition} between the low and high density phases
\emph{becomes sample-dependent} in
presence
of quenched sitewise disorder. This seems to be a property specific to
non-equilibrium systems and probably to the one dimensional case as 
the free energy of equilibrium systems (with short range
interactions) is self averaging and  the
distribution of the pseudo critical point shrinks as the system size
increases \cite{WISEMAN98}.

As in the pure case the transition is controled by the boundary parameters $\alpha$ and $\beta$, our observation that the location of the phase transition is influenced by the hopping rates on the sites close to the boundaries is not a surprise. In higher dimension, there is a large number of sites closest to the boundaries and so their effect should be averaged out.

It would be interesting to see how the other properties of the phase
diagram of the TASEP would be affected by site disorder: how the second
order transitions are modified by the disorder, how the domain wall
(uniformly distributed over the sample in the pure case \cite{APPERT02}) is distributed in the disordered case when one sits along at the first order transition point.

It would be also interesting to see whether these sample dependent phase
transitions could be seen in other non-equilibrium one dimensional
systems  \cite{KAFRI02, BARMA03} when disorder is introduced.

\end{document}